\documentclass{iau} 
\usepackage{graphicx}
\usepackage{color}

\font\fiverm=cmr5

          \font\sixrm=cmr6       

\def\teq#1{$\, #1\,$}                         

\def\actionitem#1{\textcolor{blue}{#1}}  

\def\wcyc{\omega_{\hbox{\fiverm B}}}
\def\rns{R_{\hbox{\sixrm NS}}}
\def\taueff{\tau_{{\hbox{\sixrm eff}}}}
\def\Nrec{{\cal N}_{\rm rec}}

\title[Polarized Soft X rays from Highly Magnetic Neutron Stars] 
{Polarized Radiation Signals from Highly Magnetized Neutron Star Surfaces}

\author[]   
{Kun Hu$^1$, Matthew G. Baring$^1$, Joseph A. Barchas$^2$
 \and George~Younes$^3$}

\affiliation{$^1$ Department of Physics and Astronomy - MS 108, Rice University,\\ 6100 Main St., Houston, TX 77251-1892, USA\\
$^2$ Natural Sciences, Southwest Campus, Houston Community College, \\5601 W. Loop S., Houston, Texas 77081, USA\\
$^3$ Astrophysics Science Division, NASA Goddard Space Flight Center, \\Greenbelt, MD 20771, USA}

\pubyear{2022}
\volume{363}  
\jname{Neutron Star Astrophysics at the Crossroads: Magnetars and the Multimessenger Revolution}
\editors{E. Troja \& M. G. Baring} 

\begin{document}

\maketitle

\begin{abstract}
The surfaces of neutron stars are likely sources of strongly polarized soft X rays due
to the presence of strong magnetic fields. Scattering transport in the surface layers is
critical to the determination of the emergent anisotropy of light intensity, and is
strongly influenced by the complicated interplay between linear and circular
polarization information.  We have developed a magnetic Thomson scattering simulation to
model the outer layers of fully-ionized atmospheres in such compact objects. Here we
summarize emergent intensities and polarizations from extended atmospheric simulations,
spanning considerable ranges of magnetic colatitudes.  General relativistic propagation
of light from the surface to infinity is fully included. The net polarization degrees
are moderate and not very small when summing over a variety of field directions. These
results provide an important foundation for observations of magnetars to be acquired by
NASA's new IXPE X-ray polarimeter and future X-ray polarimetry missions.

\end{abstract}

\firstsection 
\section{Introduction}

The soft X-ray thermal emission from the surfaces of neutron stars provides paths to
understanding their surfaces and interiors. Pulse profiles of isolated neutron stars
have been used to constrain the geometric parameters for different types of neutron
stars; e.g., see \cite[Gotthelf et al. (2010)]{Gotthelf10} for an example of the central
compact object PSR J0821-4300 and \cite[Younes et al. (2020)]{Younes-2020-ApJ} for the
magnetar 1RXS J170849.0–400910. Accurate interpretations of the observed surface
emission require a sophisticated understanding of the surface layer of neutron stars.
Fully or partially-ionized atmospheres of neutron stars have been extensively studied by
many groups (\cite[Shibanov et al. 1992]{Shibanov92}, \cite[Pavlov et al.
1994]{Pavlov94}, \cite[Potekhin et al. 2004]{Potekhin04}). Ionized atmosphere models in
the magnetar domain were constructed by \cite{Ozel-2001-ApJ} and \cite{HoLai-2001-MNRAS}
and several later papers. In addition, thermal emission from condensed neutron star
surfaces at relatively low temperatures has been studied by \cite[Medin \& Lai
(2006)]{Medin06} and \cite[Medin \& Lai (2007)]{Medin07}. A common feature of these and
a number of other previous studies on neutron star atmospheres is that they solve the
radiative transfer equations in terms of two orthogonal polarization modes.

In this paper, we present sample results from our Monte Carlo simulation {\sl
MAGTHOMSCATT} that treats the radiation transport in scattering-dominated neutron star
atmospheres. It can be applied to neutron stars with a wide range of magnetizations,
spanning CCO pulsars to magnetars. Our results provide a more refined determination of
the intensity and polarization signatures from the neutron star atmospheres at different
surface locales, helping to improve the interpretation of neutron star pulse profiles
and future X-ray polarization observations.  See \cite[Hu et al. (2022)]{Hu22} for a
full exposition.

\section{Monte Carlo Simulation of Radiative Transfer}

The {\sl MAGTHOMSCATT} simulation treats the radiative transfer of polarized soft X-ray 
photons in the magnetic Thomson domain (\cite[e.g., Herold 1979]{Herold79}).
It is suitably applied to the outer atmospheres of neutron stars where scatterings 
generally dominate the opacity.  The simulation code adopts a complex electric field vector 
approach, which is numerically efficient when tracking photon polarizations during 
atmospheric radiative transfer and also for general relativistic propagation through the 
magnetosphere.  The electric field vectors encapsulate full polarization information and 
capture the profound interplay between linear and circular polarizations during scatterings.  
Stokes parameters are generated at the output interface when collecting photons at infinity.

In the simulation, photons are injected at the base of the atmosphere with specified 
polarization and anisotropy, i.e. Stokes parameters.  Our adopted injection captures the 
precise polarization and anisotropy information from the high opacity solutions of radiation 
transfer due to Thomson scattering deep inside the atmosphere. Accordingly, thick 
atmospheres can be simulated efficiently with slabs of moderate opacities, \teq{\taueff =5-50}.
The design and validation of the {\sl MAGTHOMSCATT}  code is detailed in 
\cite[Barchas, Hu \& Baring (2021)]{Barchas-2021-MNRAS} and \cite[Hu et al. (2022)]{Hu22}.

\section{Polarization from Localized and Extended Atmospheres}

\begin{figure}[b]
\centerline{
 \includegraphics[width=2.6in]{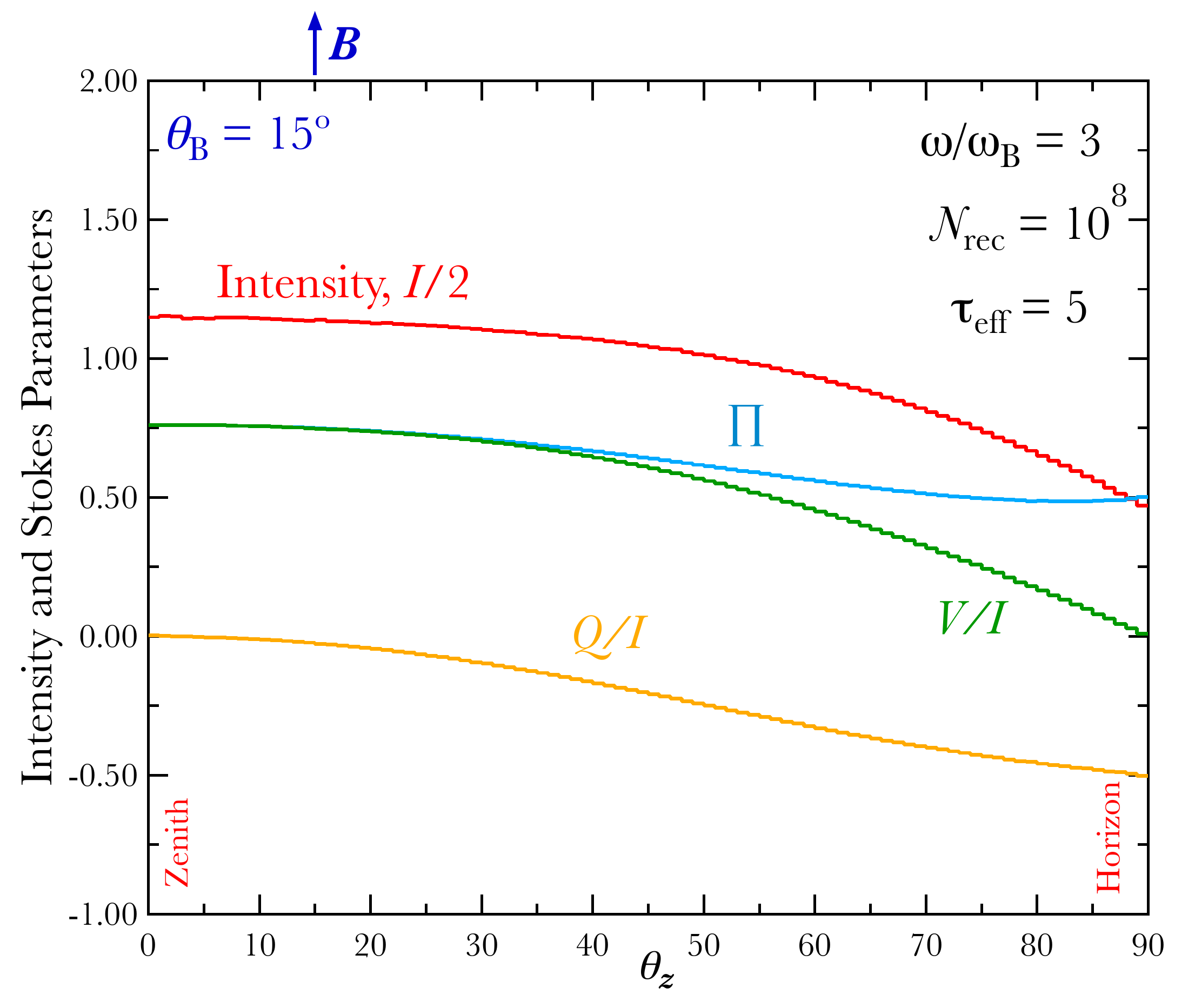} 
  \includegraphics[width=2.6in]{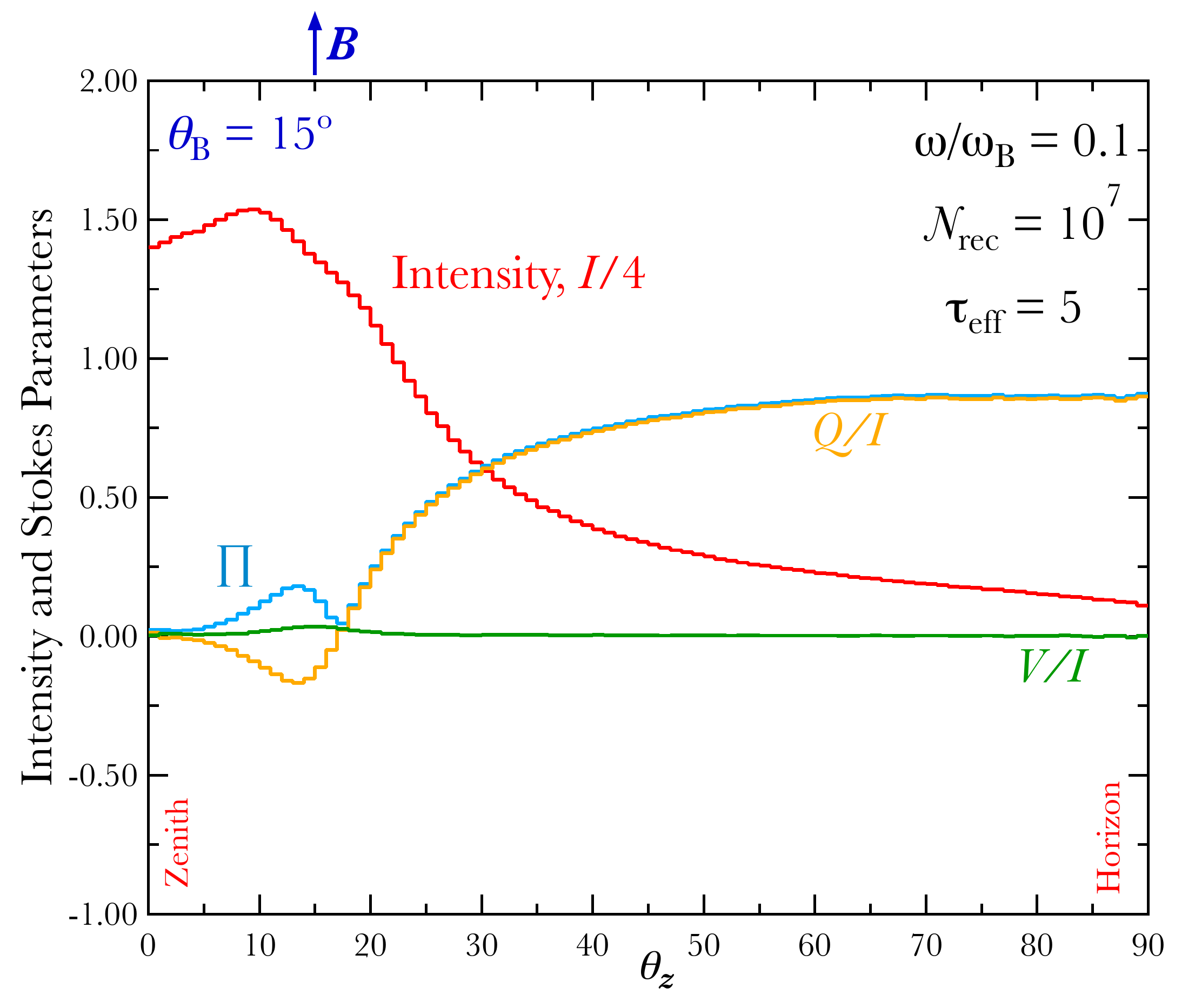} 
 }
\vspace{-5pt}
\caption{
Angular distributions of the intensity \teq{I}, Stokes \teq{Q/I}, Stokes \teq{V/I} 
and total polarization degree \teq{\Pi}, as functions of zenith angle \teq{\theta_z},  
for an atmosphere with {\bf B} inclined at \teq{\theta_{\rm B}=15^{\circ}} to the zenith.
They are for monochromatic photons 
with \teq{\omega/\wcyc = 3} (left) and \teq{\omega/\wcyc = 0.1} (right).
The effective optical depth for unpolarized photons is \teq{\taueff =5}, 
(see \cite[Hu et al. 2022]{Hu22} for the definition) and the total number of 
recorded photons \teq{\Nrec=10^7,10^8}, as indicated.  The intensities 
are scaled by factors of \teq{1/2} (left) and \teq{1/4} (right) to aid clarity of their depiction.
}
   \label{fig1}
\end{figure}

It is insightful to explore how the emergent anisotropy and polarization depend on the
photon frequency for localized atmospheric slabs. Fig.\,\ref{fig1} displays the
intensity and polarization information for monochromatic radiation emergent from the top
of slabs, in the local inertial frame at the stellar surface. 
In each panel, the magnetic field is tilted at an angle 
$\theta_{\rm B} = 15^\circ$ to the local zenith. 
This choice is germane to neutron star
emission models where radiation from hot polar caps dominates the observed flux. The
photon frequencies \teq{\omega} are scaled by the cyclotron frequency $\wcyc = eB/m_ec $. The
intensity $I$, Stokes parameters $Q$, $V$ and total polarization degree \teq{\Pi =
\sqrt{Q^2+U^2+V^2}/I} are integrated over the azimuthal angles about the local zenith
and are plotted as functions of the zenith angle $\theta_z$ of the emergent radiation.
The left panel of Fig.\,\ref{fig1} presents results at \teq{\omega/\wcyc=3}. This
approximates the ``low-field" \teq{\omega\gg \wcyc} domain, in which case the field has
limited impact on the scattering process and associated diffusion. The intensity
distribution manifests a moderate anisotropy. Linear \teq{\perp} mode polarization
prevails \teq{(Q/I<0)} when the zenith angle is not too small. The Stokes \teq{V} is
strongest along the zenith direction, reflecting the gyrational motion induced in ``radiating'' 
electrons during scatterings. The left panel of Fig.\,\ref{fig1} presents results at
\teq{\omega/\wcyc=0.1}, i.e., the strong field domain \teq{(\omega\ll \wcyc)} where
magnetic field strongly affects the scattering process. In this case, the intensity is
strongly beamed at small zenith angles close to the magnetic field direction, and \teq{\parallel} mode
photons dominate \teq{(Q/I>0)} the emergent radiation, a consequence of the character of
the cross section: see \cite[Barchas, Hu \& Baring (2021)]{Barchas-2021-MNRAS}.

\begin{figure}
\centerline{
 \includegraphics[width=2.8in]{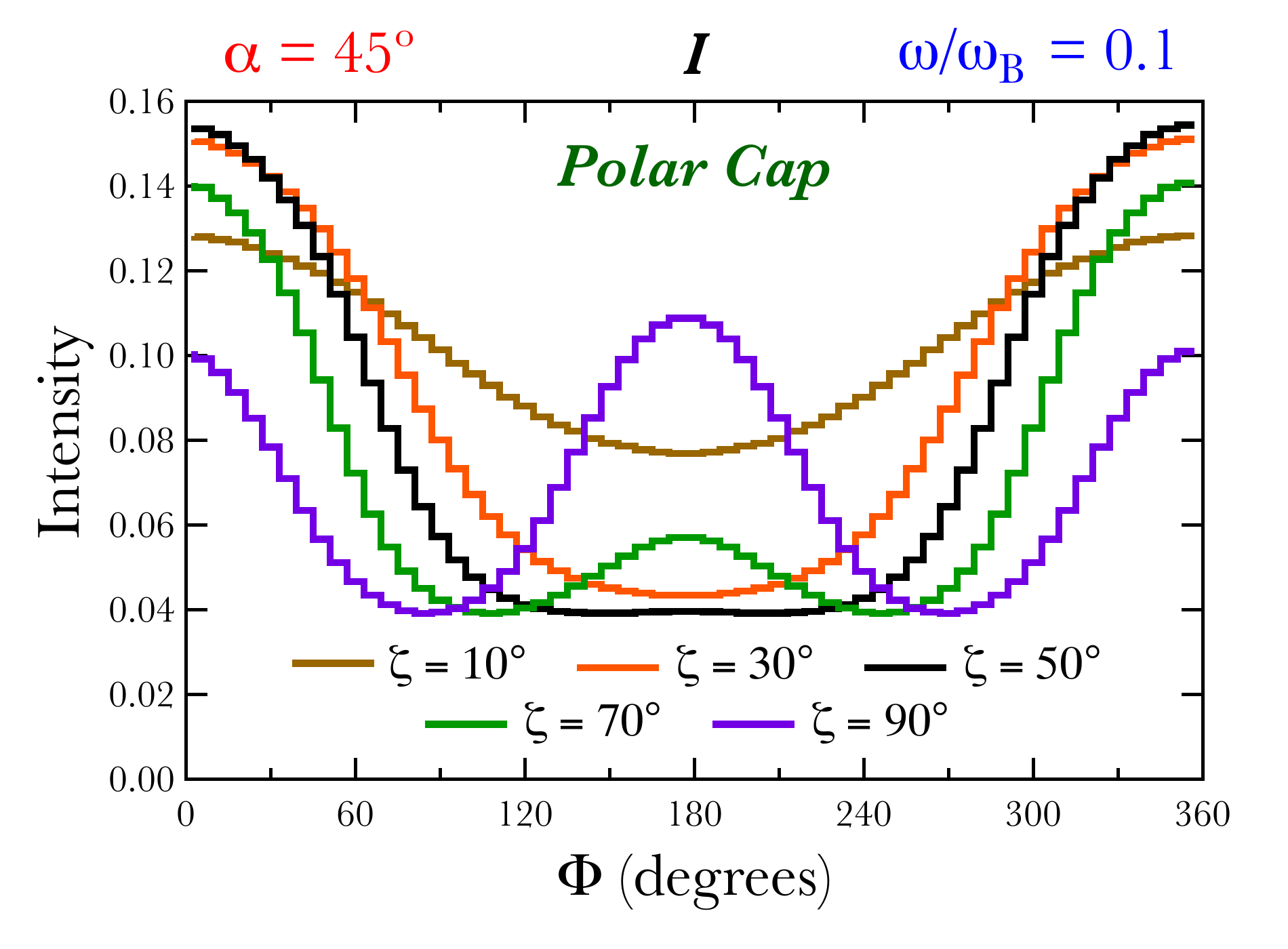} 
 \hspace{-0.5cm}
  \includegraphics[width=2.8in]{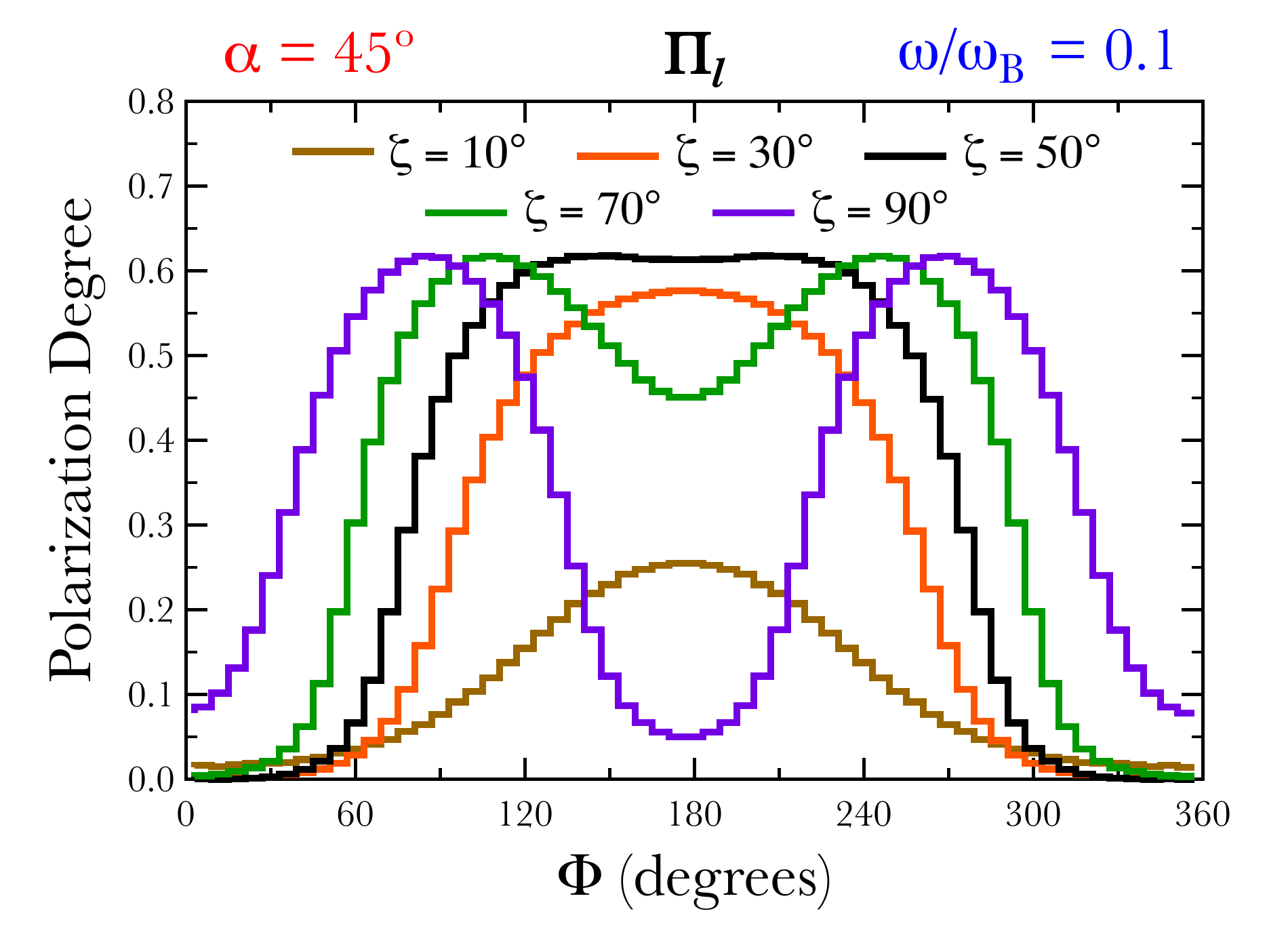} 
}
\vspace{-8pt}
\caption{
Pulse profiles of intensity \teq{I} (left) and linear polarization degree \teq{\Pi_l} (right) 
for a rotating neutron star with inclination angle \teq{\alpha = 45^{\circ}} between the 
magnetic moment and spin axes. The emission zone is a uniformly emitting polar cap 
of \teq{\theta_{\rm cap}=30^{\circ}} extent in magnetic colatitude. The profiles 
are plotted as functions of the rotational phase \teq{\Phi} 
for selected viewing angles \teq{\zeta = 10^{\circ}, 30^{\circ},} \dots \teq{90^{\circ}} to the spin axis.
The photon frequency is fixed at \teq{\omega =0.1\wcyc}.
 }
   \label{fig2}
\end{figure}

Next, we illustrate the intensity and polarization signatures from extended atmospheric
regions. This is achieved by sampling the emission locales for each photon uniformly
across the pertinent surface region, mimicking isothermal conditions. General
relativistic light bending and the parallel transport of the polarization vectors to
infinity are now treated, assuming \teq{M = 1.44M_{\odot}} and \teq{\rns = 10^6} cm for
the star. In Fig.\,\ref{fig2}, the simulated intensity \teq{I} and \underline{linear}
polarization degree \teq{\Pi_l=\sqrt{Q^2+U^2}/I} are plotted as functions of rotational
phase \teq{\Phi} for different viewing directions \teq{\zeta} and a fixed inclination
angle \teq{\alpha=45^{\circ}}; both angles are measured relative to the stellar rotation
axis. The emission regions are chosen to be two antipodal polar caps with half-opening
angles equal to \teq{30^{\circ}}, and the photon frequency ratio is set to
\teq{\omega/\wcyc=0.1}. The intensity profiles demonstrate a strong rotational
modulation with a pulse fraction \teq{\lesssim 50\%}. The maximum linear polarization
degree \teq{\Pi_l} is around \teq{60\%}. This \teq{\Pi_l} is very high because the
depolarization induced by sampling photons from emission locales with different field
directions is reduced when restricting locales to polar caps. This high polarization
degree is of interest to NASA's recently launched IXPE, and also future X-ray
polarimeters.

\section{Observational Diagnostic Potential}

\begin{figure}
\centerline{
 \includegraphics[width=2.8in]{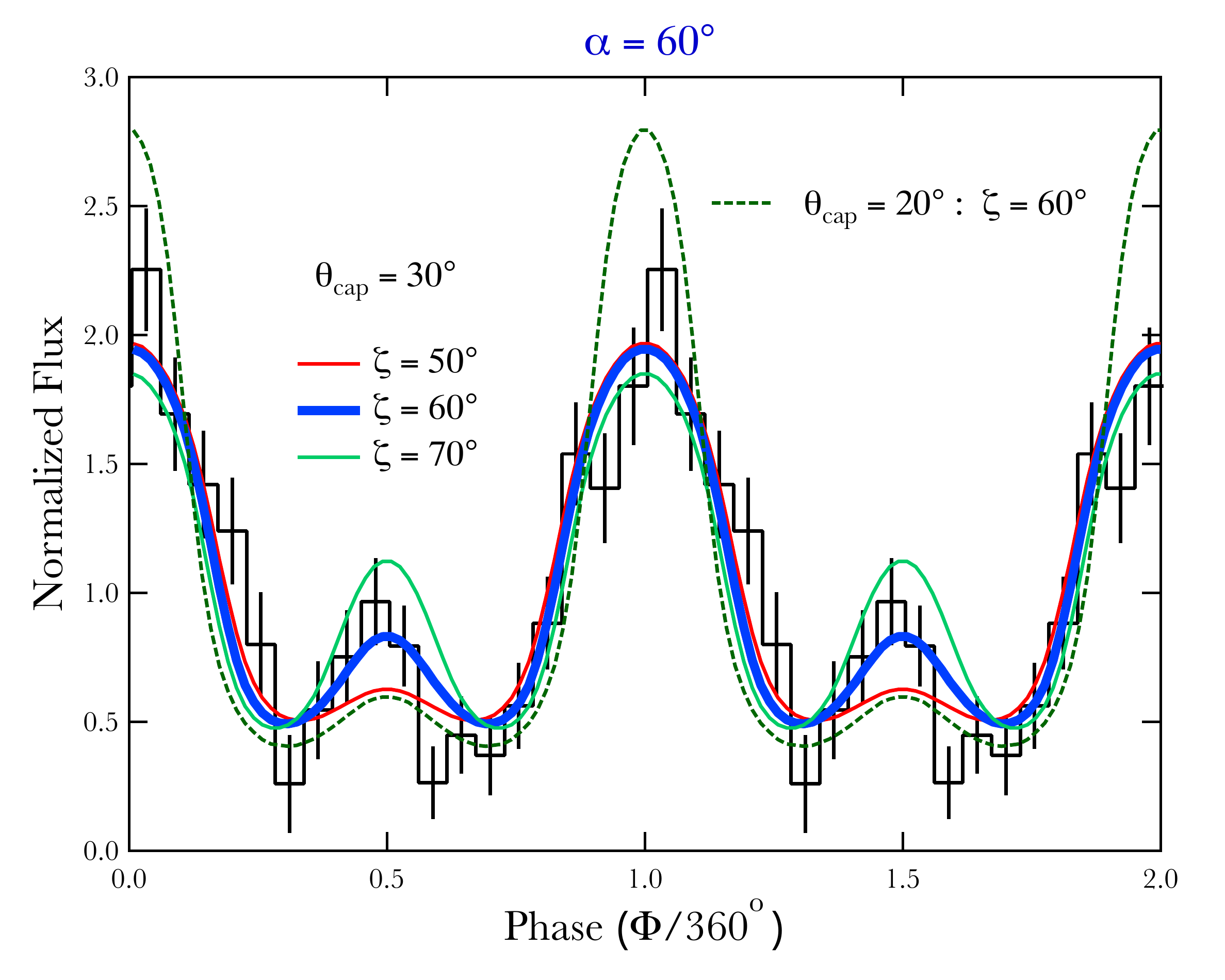} 
  \hspace{-0.5cm}
  \includegraphics[width=2.8in]{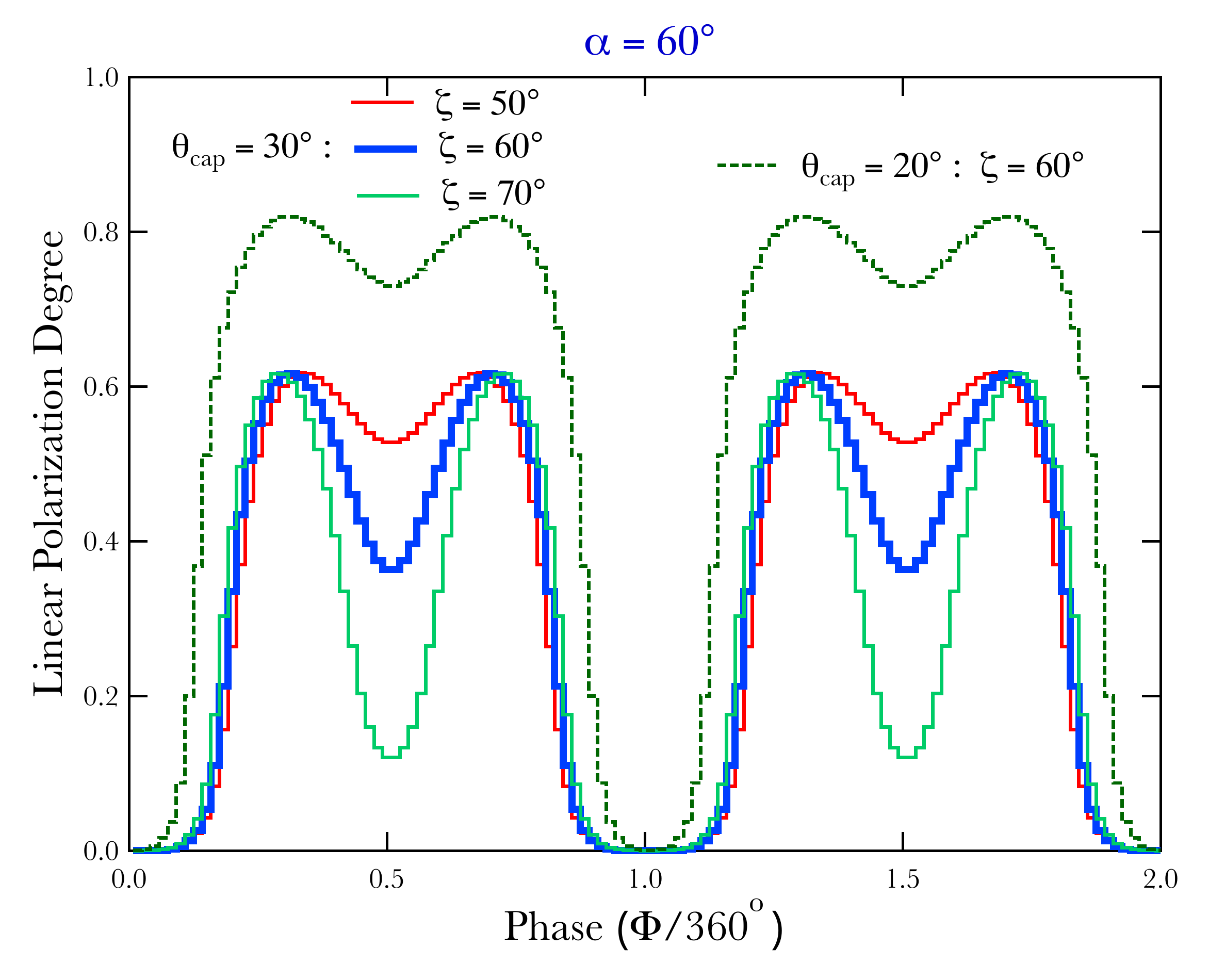} 
 }
\vspace{-8pt}
 \caption{
 \actionitem{Left:}
    Simulated pulse profiles for intensity \teq{I} for two antipodal 
    polar caps extending from the respective magnetic poles to colatitudes 
    of \teq{\theta_{\rm cap}=30^{\circ}}.  
    The histogram represents the data in Fig.~3 (right) of \cite{Younes-2020-ApJ} for 
    the flare contribution to persistent emission of 1RXS J1708-40.  The theoretical 
    models center on the pulse profile (heavyweight, blue) for \teq{\alpha = 60^{\circ}
    = \zeta}, which was the system configuration that gave the best 
    statistical fit to the observations.  The other two solid curves 
    are for models in the parametric neighborhood with \teq{\zeta}
    values as labeled.  The dotted green curve is for a smaller polar cap with \teq{\theta_{\rm cap}=20^{\circ}}.  
  \actionitem{Right:}
    The linear polarization \teq{\Pi_l} pulsation traces corresponding to the 
    intensity curves on the left. }
   \label{fig3}
\end{figure}

Given the strong sensitivity of emission to the stellar geometric angles \teq{\alpha},
\teq{\zeta}, and the surface distribution of emission locales, we can compare
predictions from {\sl MAGTHOMSCATT} with observed intensity pulse profiles. We performed
an exercise of constraining the geometric parameters of the magnetar 1RXS J1708-40.
Intensity pulse profiles for models with different geometric parameters are plotted in
the left panel of Fig.\,\ref{fig3}, with the best-fit one displayed as blue, indicating
a preferred viewing angle \teq{\zeta \sim 60^{\circ}} to the spin axis.  The choice of
\teq{\omega = 0.1\wcyc} was made for expediency, generating good statistics with
reasonable run times.  Since the dominant contribution to the signal comes from
\teq{\sim 15^{\circ}-30^{\circ}} colatitudes, the typical emergent photon angles sampled
relative to the local {\bf B} direction are large enough that the cross section is
somewhat insensitive to the choice of \teq{\omega}, obviating the need to treat slow
\teq{\omega\sim 10^{-4}\wcyc} runs. The right panel presents the pulse profiles of
linear polarization degrees for selected models. Refined probes of the geometry
parameters can be acquired once phase-resolved polarization observations are available:
for brighter magnetars, IXPE is likely to deliver such during the coming year.

\end{document}